\newcommand{\co}[2]{\ifcase #1  \or #2 \fi}
\newcommand{\unit}[1]{\,{\rm #1}}

\documentclass[prb,showpacs,twocolumn,byrevtex,showkeys,floatfix]{revtex4}

\usepackage{graphicx}
\usepackage{color}
\usepackage{dcolumn}
\usepackage{amsmath}
\usepackage{amssymb}
\usepackage[colorlinks=false]{hyperref}

\include{MyMc} 
\begin{document}

\title{
  Non-ideal artificial phase discontinuity in long Josephson 0-$\kappa$-junctions
}

\author{T.~Gaber}
\email{gaber@uni-tuebingen.de}
\author{E.~Goldobin}
\author{A.~Sterck}
\author{R.~Kleiner}
\author{D.~Koelle}
\affiliation{
  Physikalisches Institut, Experimentalphysik II,
  Universit\"at T\"ubingen,
  Auf der Morgenstelle 14,
  D-72076 T\"ubingen, Germany
}
\author{M.~Siegel}
\author{M.~Neuhaus}
\affiliation{
 Universit\"at Karlsruhe,
 Institut f\"ur Mikro-- und Nanoelektronische Systeme,
 Hertzstr. 16,
 D-76187 Karlsruhe, Germany
}


\date{\today}

\begin{abstract}

We investigate the creation of an arbitrary $\kappa$-discontinuity
of the Josephson phase in a long Nb-AlO$_x$-Nb Josephson junction
(LJJ) using a pair of tiny current injectors, and study the
formation of fractional vortices formed at this discontinuity. The
current $I_{\rm inj}$, flowing from one injector to the other,
creates a phase discontinuity $\kappa\propto I_{\rm inj}$. The
calibration of injectors is discussed in detail. The small but
finite size of injectors leads to some deviations of the
properties of such a 0-$\kappa$-LJJ from the properties of a LJJ
with an ideal $\kappa$-discontinuity. These experimentally
observed deviations in the dependence of the critical current on
$I_{\rm inj}$ and magnetic field  can be well reproduced by
numerical simulation assuming a finite injector size. The physical
origin of these deviations is discussed.

\end{abstract}

\maketitle

\section{Introduction}
\label{Sec:Intro}

It has been shown theoretically and experimentally that by
utilizing unconventional superconductors or ferromagnetic barriers
one can fabricate so-called Josephson
$\pi$-junctions\cite{Bulaevskii:pi-loop,Kontos:2002:SIFS-PiJJ,Ryazanov:2001:SFS-PiJJ,Lombardi:2002:dWaveGB,VanHarlingen:1995:Review}($\pi$
JJs). While for conventional 0-junctions the first Josephson
relation reads $I_s = I_c \sin \mu$, for $\pi$-junctions it is
$I_s = -I_c \sin \mu = I_c \sin (\mu + \pi)$.

In a 1D long Josephson junction (LJJ) with alternating 0 and $\pi$
regions  vortices carrying half of a magnetic flux quantum
$\Phi_0$ (so-called semifluxons~\cite{Goldobin:SF-Shape}) can
spontaneously appear at the boundaries between $0$ and $\pi$
regions.\cite{Bulaevskii:0-pi-LJJ,Xu:SF-shape} Such semifluxons
were observed experimentally.
\cite{Hilgenkamp:zigzag:SF,Kirtley:SF:HTSGB,Kirtley:SF:T-dep,Tsuei:Review,Sugimoto:TriCrystal:SF}

In contrast to an integer Josephson vortex (fluxon), which is a
free moving soliton, a semifluxon is like a spin $\frac{1}{2}$
particle
--- it is pinned at a $0$-$\pi$ boundary, has one of two possible
polarities $\pm\frac12\Phi_0$ and, the most important, can form
the groundstate of the system. Semifluxons are very interesting
non-linear objects: they can form a variety of
groundstates,\cite{Kogan:3CrystalVortices,Goldobin:SF-ReArrange,Zenchuk:2003:AnalXover,Susanto:SF-gamma_c,Kirtley:IcH-PiLJJ}
may
flip,\cite{Hilgenkamp:zigzag:SF,Kirtley:SF:T-dep,Goldobin:F-SF}
rearrange\cite{Goldobin:SF-ReArrange} or get
depinned\cite{Susanto:SF-gamma_c,Lazarides:Ic(H):SF-Gen,Stefanakis:ZFS/2}
by a bias current, lead to half-integer zero-field
steps\cite{Stefanakis:ZFS/2,Lazarides:Ic(H):SF-Gen,Goldobin:Art-0-pi}
and have a characteristic
eigenfrequency.\cite{Goldobin:2KappaEigenModes} Huge arrays of
fractional flux
quanta\cite{Hilgenkamp:zigzag:SF,Zenchuk:2003:AnalXover,Goldobin:SF-ReArrange,Susanto:SF-gamma_c}
were realized and predicted to have a tunable plasmon band
structure\cite{Susanto:1D-Crystal} which can be thought of as a
plasmonic crystal similar to photonic crystals. Semifluxons are
also promising candidates for storage devices in classical or
quantum domain. They have a very small effective mass\endnote{In
this context "effective mass" actually means momentum of inertia
$\sim 10^{-3}m_e\lambda^{2}_J$ ($m_e$ is the electron mass and
$\lambda_J$ the Josephson penetration
depth)\cite{Goldobin:QuTu2Semifluxons}} and
the potential for the observation of macroscopic quantum effects and building qubits. 

Recently we have demonstrated that one can also study semifluxons
in a conventional Nb-AlO$_x$-Nb LJJ.\cite{Goldobin:Art-0-pi} This
is done by creating a discontinuity of the Josephson phase $\phi$
(the difference between $\phi$ and $\mu$ is explained below) by
means of current injectors.
\begin{figure}[!tb]
  \centering \includegraphics*{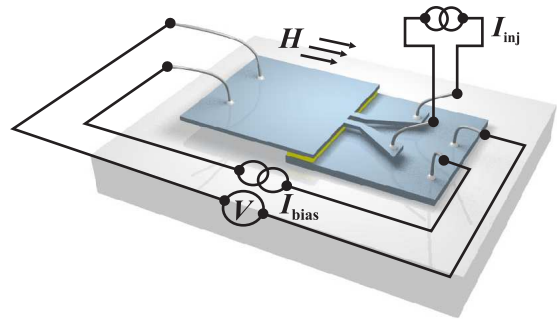}
  \caption{Sketch of a Josephson junction with a pair of current
  injectors to create an arbitrary discontinuity of the Josephson
  phase.
  }
  \label{Fig:sketch}
\end{figure}
Two injectors of width $\Delta w$ are attached to one of the
junction's electrodes at a distance $\Delta x$ from each other, as
shown in Fig.~\ref{Fig:sketch}. To create an ideal discontinuity,
both the injector width $\Dw$ and the distance $\Dx$ between them
must be much smaller than the Josephson penetration depth
$\lambda_J$. The current $I_{\rm inj}$, that is passed from one
injector to another creates a phase shift $\kappa\propto I_{\rm
inj}$ across the small part $\sim 2\Delta w+\Delta x$ of the top
electrode. Since $\kappa \propto I_{\rm inj}$, the phase shift
$\kappa$ can be tuned to $\kappa = \pi$ by changing $I_{\rm inj}$.
When the discontinuity of the phase is created, the fractional
vortex centered at the discontinuity immediately appears to
compensate it.

LJJs with such artificially created discontinuities have a number
of advantages. First of all, by creating a discontinuity
$\kappa\ne\pi$ one can study not only semifluxons, but vortices
with any \emph{arbitrary} flux.\cite{Goldobin:2KappaGroundStates}
The variable $\kappa$ can also be used as tuning parameter in some
devices, \eg, to tune the plasmon band
structure.\cite{Susanto:1D-Crystal} Second, due to low damping in
Nb-AlO$_x$-Nb LJJs one can study the \emph{dynamics} of the
fractional vortices. Exponentially low damping at $T \ll T_c$ due
to the energy gap also helps to build qubits with good decoherence
figures.

LJJs with double injector structure can also be used for the
controllable insertion of an integer fluxon ($\kappa=2\pi$) into
the
LJJ,\cite{Ustinov:2002:ALJJ:InsFluxon,Malomed:2004:ALJJ:Ic(Iinj)}
\eg, to set/read out the state of a fluxon or semifluxon based
(qu)bit.\cite{Goldobin:F-SF,Kemp:2002:JVQ-Readout,Kaplunenko:2004:VortQubitIface}
Thus, the structure shown in Fig.~\ref{Fig:sketch} can be used in
a range of the electronic devices as well as for basic studies
involving integer and fractional Josephson vortices.

In experiments, one of the main steps is the calibration of the
injectors, \ie, the experimental determination of the ratio
$\kappa/I_{\rm inj}$. Although a direct theoretical calculation of
the ratio $\kappa/I_{\rm inj}$ is possible in the framework of a
1D model presented below, in practice a different approach is
easier and more reliable: one measures the $I_c(I_{\rm inj})$
dependence and compares it with the $I_c(\kappa)$ dependence
obtained theoretically.

The purpose of this paper is to present this calibration method,
\ie, study how the $I_c(\kappa)$ and, therefore, the $I_c(I_{\rm
inj})$ dependence should look like for typical experimental
parameters. We especially are interested in the effects caused by
the finite injector sizes, \ie, when the discontinuity is not an
ideal step function. 

This paper is organized as follows. In section \ref{Sec:Model} we
introduce the model of a LJJ with phase discontinuities. In
section~\ref{Sec:IdealDiscont} we first consider $\delta$-like
current injectors that correspond to an ideal step-like
discontinuity (or discontinuity \emph{point}) of the Josephson
phase and discuss the $I_c(\kappa)$ dependence. In
section~\ref{Sec:Samples} we compare our theory with experimental
results. As there are certain misfits between measurements and
theoretical predictions, in Sec.~\ref{Sec:IdealDipole} we take the
finite size of the injectors into account. Namely, we investigate
the effect of the finite injector size on the $I_c(\kappa)$
dependence and on the dependence of $I_c$ on magnetic field $H$
and compare them with experimental results. 
Section~\ref{Sec:Conclusion} concludes this work.

\section{The Model}
\label{Sec:Model}

The dynamics of the Josephson phase in a LJJ consisting of $0$-
and $\pi$-parts can be described by the 1D time-dependent
perturbed sine-Gordon
equation\cite{Goldobin:SF-Shape,Malomed:2004:ALJJ:Ic(Iinj),Ustinov:2002:ALJJ:InsFluxon}
\begin{equation}
  \phi_{xx}-\phi_{tt}-\sin\phi = \alpha\phi_t-\gamma(x)-\theta_{xx}(x)
  , \label{Eq:sG-phi}
\end{equation}
where $\phi(x,t)$ is the Josephson phase and subscripts $x$ and
$t$ denote the derivatives with respect to coordinate $x$ and time
$t$. In Eq.~(\ref{Eq:sG-phi}) the spatial coordinate $x$ is
normalized to the Josephson penetration depth $\lambda_J$ and the
time $t$ is normalized to the inverse Josephson plasma frequency
$\omega_p^{-1}$; $\alpha=1/\sqrt{\beta_c}$ is the dimensionless
damping parameter; $\gamma=j/j_c$ is the external bias current
density normalized to the critical current density of the
junction. The function $\theta(x)$ is a step function which is
discontinuous at the points where the $0$- and $\pi$-parts join
and constant elsewhere. For example in 0-$\pi$-LJJs, $\theta(x)$
can be equal to zero along all 0-parts and to $\pi$ along all
$\pi$-parts. For simplicity we assume that there is only one
discontinuity point at the center of the junction, $x=0$.

\begin{figure}[!tb]
  \centering \includegraphics*{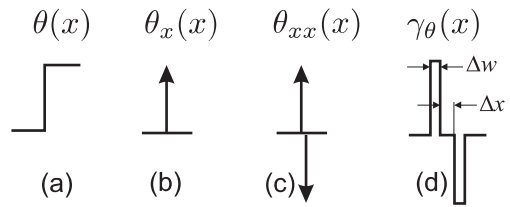} 
   \caption{
      Functions $\theta(x)$, $\theta_x(x)$, $\theta_{xx}(x)$ and
      the approximation $\gamma_\theta(x)$ of $\theta_{xx}(x)$ by two rectangular pulses
      of width $\Delta w$ and distance $\Delta x$.
  }
  \label{Fig:dipole_layout}
\end{figure}

It is clear from Eq.~(\ref{Eq:sG-phi}) that the solution $\phi(x)$
is also $\pi$-discontinuous at the same points as $\theta(x)$.
Therefore, we often call the points where 0- and $\pi$-parts join
\emph{phase
discontinuity points}.\\
The bias current density $\gamma(x)$ and
$\theta_{xx}(x)$ are two additive terms in Eq.~(\ref{Eq:sG-phi}).
This means that if one does not have initially a $\theta_{xx}(x)$
term (like in a conventional LJJ), its effect may be substituted
by an additional bias current $\gamma_{\theta}(x)=\theta_{xx}(x)$.
If $\theta(x)=\pi \Heaviside(x)$ is a step function
[Fig.~\ref{Fig:dipole_layout}(a)], where $\Heaviside(x)$ is the Heaviside
step function, then $\theta_x(x)=\pi\delta(x)$ is a
$\delta$-function of strength $\pi$ [Fig.~\ref{Fig:dipole_layout}(b)], and
$\theta_{xx}(x)=-\pi\delta(x)/x$
[Fig.~\ref{Fig:dipole_layout}(c)].
Thus the $\pi$-discontinuity of the phase corresponds to (or can
be substituted by) an additional current with the density
$-\pi\delta(x)/x$, \ie, by a \emph{current dipole} of strength
$\pi$.\cite{Malomed:2004:ALJJ:Ic(Iinj), Goldobin:Art-0-pi,
Ustinov:2002:ALJJ:InsFluxon} Of course, in experiment one cannot
pass infinitely large currents via infinitesimal wires, so the
real current profile will be an approximation of $\theta_{xx}(x)$,
\eg, like the one shown in Fig.~\ref{Fig:dipole_layout}(d).
Nevertheless, following
Ref.~\onlinecite{Malomed:2004:ALJJ:Ic(Iinj)}, we first consider an
ideal current dipole ($\delta$-like current dipole) corresponding
to an ideal discontinuity.

It is clear, that when we speak in the language of current dipoles
we can change the amplitude of the current, \ie, use
$-\kappa\delta(x)/x$, which corresponds to a
$\kappa$-discontinuity of the Josephson phase (instead of a
$\pi$-discontinuity). Below we consider a LJJ with such an
arbitrary $\kappa$-discontinuity of the Josephson phase, \ie,
$\theta(x)=\kappa \Heaviside(x)$ in
Eq.~(\ref{Eq:sG-phi}).\\
Note that other
authors\cite{Xu:SF-shape,Kirtley:IcH-PiLJJ,Buzdin:2003:phi-LJJ}
often use directly the equation written for the {\it continuous}
phase $\mu(x,t)$, which for the case of a $\kappa$-discontinuity
reads
\begin{equation}
  \mu_{xx}-\mu_{tt}\sin \left( \mu+\theta(x) \right) = \alpha\mu_t-\gamma(x)
  . \label{Eq:sG-mu}
\end{equation}
Equation~(\ref{Eq:sG-phi}) and (\ref{Eq:sG-mu}) are, actually,
equivalent and one can be obtained from the other by substitution
$\phi(x,t)=\mu(x,t)+\theta(x)$.\cite{Goldobin:SF-Shape} To
simulate the system we used the {\sc ``StkJJ''}
software,\cite{StkJJ} which implements finite difference algorithms for solving Eq.~(\ref{Eq:sG-mu}).\\

\section{Ideal Discontinuity}
\label{Sec:IdealDiscont}
Using Eq.~(\ref{Eq:sG-phi}) we first would like to know how the
normalized maximum supercurrent  \[
i_c=\frac{I_c(\kappa,H)}{I_c(0,0)} \] depends on $\kappa$. Here
$i_c$ can be obtained by varying the normalized supercurrent
\[ i=\frac{1}{l} \int_{-l/2}^{l/2} \sin\!\phi\,dx \] with respect to
$\phi$, which should still be a solution of Eq.~(\ref{Eq:sG-phi}).
The length $l=L/\lambda_J$ is the normalized LJJ length. So far
the $i_c(\kappa)$ dependence was calculated only for an annular
LJJ.\cite{Malomed:2004:ALJJ:Ic(Iinj),Nappi:2002:ALJJ-Ic(I_inj)} It
was predicted and measured\cite{Ustinov:2002:ALJJ:InsFluxon} that
this dependence is a Fraunhofer pattern with the first minimum at
$\kappa=2\pi$. This result \emph{does not depend} on the junction
length $L$.

In a junction of linear geometry the situation is different and
$i_c(\kappa)$ \emph{depends} on the normalized length $l$. Let us
first consider extreme cases.

\begin{figure}[!tb]
  \centering\includegraphics*{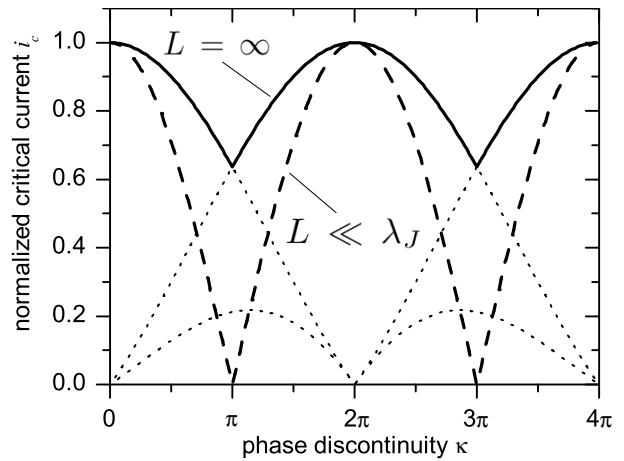}
  \caption{
      Dependencies of the normalized critical current $i_c$ on the value of
      the Josephson phase discontinuity $\kappa$. Dotted lines show
      Fraunhofer patterns (\ref{Eq:Fraunhofer}) shifted by $0,2\pi,4\pi$,  the corresponding $i_c(\kappa)$
      for $L=\infty$ is shown by solid lines. Dashed lines show the dependence (\ref{Eq:cos})
      for $L\ll\lambda_J$.
  }
  \label{Fig:Ic(kappa)}
\end{figure}

1. $l \gg 1$. In this case one can follow the derivation of
Ref.~\onlinecite{Malomed:2004:ALJJ:Ic(Iinj)} and arrive to the
same Fraunhofer pattern (cf. Eq.~(10) of
Ref.~\onlinecite{Malomed:2004:ALJJ:Ic(Iinj)})
\begin{equation}
  i_c(\kappa)=
  \left| \frac{\sin(\kappa/2)}{\kappa/2} \right|
  . \label{Eq:Fraunhofer}
\end{equation}
On the other hand it is clear, that a $\kappa$-discontinuity and a
$(\kappa+2\pi)$-discontinuity in linear LJJs are physically
equivalent, since the phase is defined modulo $2\pi$ and there are
no topological limits as in annular LJJs. Therefore, the
Fraunhofer pattern (\ref{Eq:Fraunhofer}) should be replicated
periodically with a period of $2\pi$ along the $\kappa$-axis, as
shown in Fig.~\ref{Fig:Ic(kappa)} (dotted lines). Hence, for a
given value of $\kappa$ one may have more than one $i_c$,
corresponding to different solutions. If one sweeps $\kappa$ at
fixed $i$, or sweeps $i$ at fixed $\kappa$ the system may switch
from one solution to another by emitting a fluxon which leaves the
junction and does not affect the system later. This happens when
the bias point crosses one of the dotted lines in
Fig.~\ref{Fig:Ic(kappa)}. We note that in an annular LJJ a fluxon
cannot escape. Therefore transitions as described above (for a
linear LJJ) cannot occur, and only the principal Fraunhofer curve
Eq.~(\ref{Eq:Fraunhofer}) has to be considered.
We do not discuss the stability of each solution here, but it is
important to note that the topmost branch, indicated by the solid
line in Fig.~\ref{Fig:Ic(kappa)}, separates all Meissner states
from the finite voltage state and can be measured in experiment.
Thus, one gets cusp-like minima at $\kappa=\pi+2\pi n$ (with
integer $n$).

2. $l\ll 1$. In case of a short junction one can neglect the
spatial variation of the Josephson phase along the junction and
assume that the phase in the left half of the junction is $\phi_0$
and in the right half of the junction is $\phi_0+\kappa$. By
varying $\phi_0$ we find that the critical current depends on
$\kappa$ as
\begin{equation}
  i_c(\kappa)=
  \left|\,\cos\left( \frac{\kappa}{2} \right) \right|
  , \label{Eq:cos}
\end{equation}
which is shown in Fig.~\ref{Fig:Ic(kappa)} by the dashed line.

In Fig.~\ref{Fig:Ic(kappa)} one can see that both dependences have
a common feature: they have maxima at $\kappa=2\pi n$ and
cusp-like minima (possibly with hysteresis) at $\kappa=\pi+2\pi
n$. The value of $i_c$ at the minima depends on the junction
length and varies from 0 for a short junction to $2/\pi$ for an
infinitely long one.

3. Intermediate $l$. In this case the $i_c(\kappa)$ curve lies in
between the two curves for $l \ll 1$ and $l \gg 1$, but still has
cusp-like minima at $\kappa=\pi(2n+1)$. The numerically calculated
dependence of the minimum critical current $i_c(\pi)$ on junction
length will be
shown later in Sec.~\ref{Sec:IdealDipole}. \\
Since cusp-like minima in $i_c(\kappa)$ always correspond to a
phase discontinuity $\kappa=(2n+1)\pi$, the calibration process of
the injectors is simple. One should measure $I_c(I_{\rm inj})$.
Since $\kappa \propto I_{\rm inj}$ the value of $I_{\rm inj}$ at
the first cusp-like minimum of the $I_c(I_{\rm inj})$ dependence
corresponds to $\kappa$ mod $2\pi = \pi$ .

\section{Samples}
\label{Sec:Samples}

So far an ideal $\kappa$-discontinuity or, to put it in terms of
an additional bias current, $\delta$-like current injectors were
discussed and the $i_c(\kappa)$ dependence was investigated. As a
result we saw that the $i_c(\kappa)$ curve is 2$\pi$ periodic and
has a minimum of the critical current for phase discontinuities
$\kappa=(2n+1)\pi$ independent of the junction length.\\ In order
to experimentally investigate artificial $0$-$\kappa$-LJJs we used
samples that were fabricated at the University of Karlsruhe as
well as by Hypres\footnote{Hypres, Elmsford, New York, USA.
http://www.hypres.com} and that are based on conventional
Nb-AlO$_x$-Nb technology. All data presented here have been
obtained at $T = 4.2\,{\rm K}$ from different samples with
parameters summarized in Tab.~\ref{Tab:Samples}.
\begin{table*}
\centering
\begin{tabular}{|c|>{\centering}m{1.35cm}|>{\centering}m{0.9cm}|>{\centering}m{1.50cm}|>{\centering}m{1.5cm}|>{\centering}m{2cm}|c|>{\centering}m{2.4cm}|c|c|} \hline
   chip & $j_c$ [A/cm$^2$]& $\lambda_J$ [$\mu$m] & length $L$ \newline [$\mu$m] & norm. \newline length $l$ & injector \newline width [$\mu$m]& $\Delta w$ & injector \newline distance [$\mu$m]& $\Delta x$ &
   $N$ \\ \hline
   \#1\footnotemark[1] \footnotetext[1]{fabricated at the University of Karlsruhe, Germany.} & $\approx$ 150 & $\approx$ 24 & 30--360 & 1.25--15 & 5 & 0.21 & 5 & 0.21 & 7\\
    \#2\footnotemark[1] & $\approx$ 400 & $\approx$ 15 & 60 & 4 & 2 & 0.13 & 2 & 0.13 & 1\\
    \#3\footnotemark[1] & $\approx$ 400 & $\approx$ 15 & 120 & 8 & 5 & 0.33 & 5 & 0.33 & 1\\
     \#4\footnotemark[2] \footnotetext[2]{fabricated by Hypres, Elmsford, New York, USA.}& $\approx$ 100 & $\approx$ 30 & 60 & 2 & 5 & 0.17 & 4 & 0.13 & 2\\ \hline
\end{tabular}
  \caption{Sample properties. The quantities $l$, $\Delta w$ and $\Delta x$ are the normalized (with respect to $\lambda_J$) LJJ length, injector width and distance between the injectors. $N$ is the number of junctions on the chip, that were measured. The junction width is $5\unit{\mu m}$ for all samples.}
\label{Tab:Samples}
\end{table*}
For all samples the junction layout is identical to the one
presented in Fig.~\ref{Fig:sketch}. A photograph can be found in
Fig.~3 of Ref.~\onlinecite{Goldobin:Art-0-pi} (sample~\#4 in
Tab.~\ref{Tab:Samples}). However, the values of $\lambda_J$, the
lengths $L$ and the injector sizes are different. For all samples
the $I$-$V$ characteristics (IVC) and the dependences of $I_c(H)$
without injector current were measured to
ensure good sample quality and the absence of trapped magnetic flux (not shown). \\
\begin{figure}[!tb]
  \centering \includegraphics*{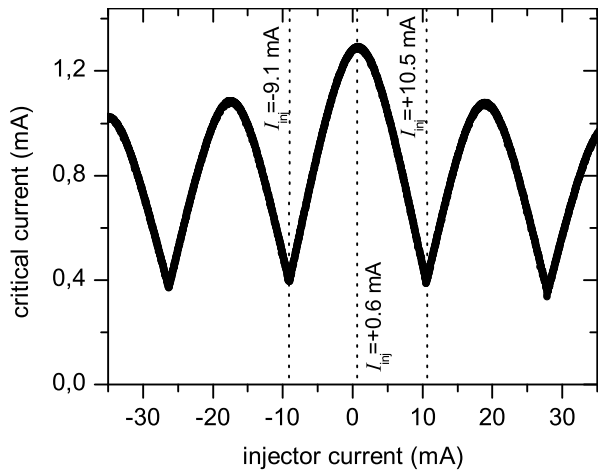} 
  \caption{Experimentally measured $I_c(I_{\rm inj})$ dependence
  at $H=0$ (sample \#2). 
  }
  \label{Fig:Ic(inj)_measured}
\end{figure}
Figure~\ref{Fig:Ic(inj)_measured} shows a typical measurement of
$I_c(I_{\rm inj})$ at $H=0$, obtained from sample~\#2. One can
clearly see the almost periodic modulation of the critical current
as a function of $I_{\rm inj}$ which is very similar to the one
predicted in Fig.~\ref{Fig:Ic(kappa)}. Yet, there is an unexpected
decrease of $I_c$ at the side maxima $I_{\rm inj} \approx \pm
19\unit{mA}$. We have also measured the $I_c(H)$ dependence at
$I_{\rm inj}$ corresponding to the first minimum in the
$I_c(I_{\rm inj})$ dependence, \ie, the $I_c(H)$ dependence in the
0-$\pi$-state. Such a dependence is expected to have a minimum of
$I_c$ at $H=0$.\cite{Kirtley:IcH-PiLJJ,Smilde:ZigzagPRL}
\begin{figure*}[!tb]
  \centering \includegraphics* {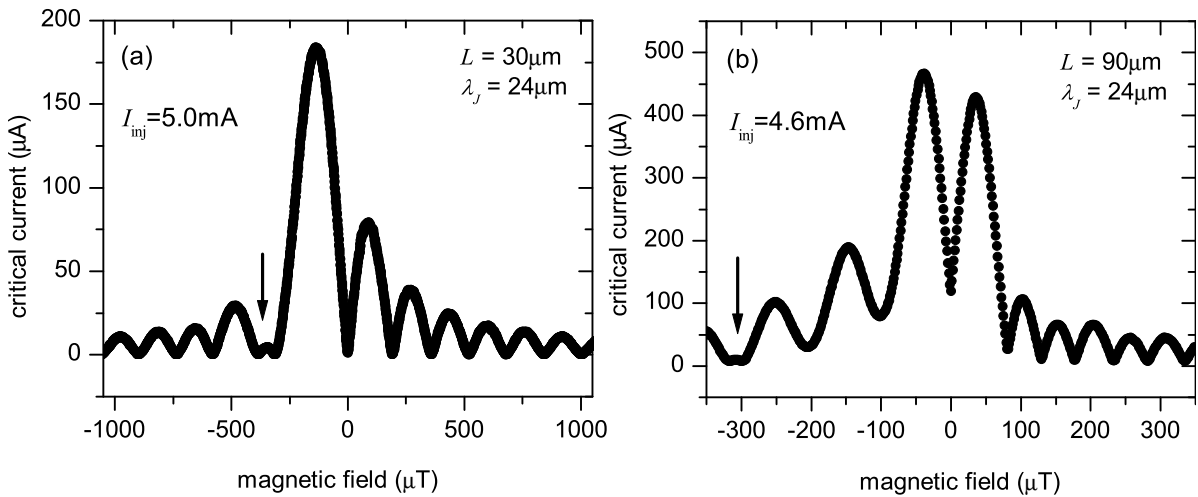}
  \caption{Measured $I_c(H)$ dependences in the 0-$\pi$-state for two LJJ (chip \#1) of length $L =
  1.25\,\lambda_J$ (a) and $L = 3.75\,\lambda_J$ (b).
  Arrows mark the crossing of the $H$-axis in negative fields [cf. Fig.~\ref{Fig:Ic(H)_maplesim} (b)].}
  \label{Fig:Ic(H)_measured}
\end{figure*}
In Fig.~\ref{Fig:Ic(H)_measured} two typical $I_c(H)$ dependences
experimentally obtained from different samples are presented.
Although we indeed observe a minimum of $I_c$ at
$H=0$,\cite{VanHarlingen:1995:Review,Smilde:ZigzagPRL} the curves
in Fig.~\ref{Fig:Ic(H)_measured} are rather asymmetric with
respect to $H=0$:\cite{Goldobin:Art-0-pi} the amplitudes of the
first maxima are different and the higher order maxima have
different, sometimes not very regular, periods.

In the following we show that these deviations can be reproduced
in numerical simulations using the model with the finite size of
the injectors. The physical origin of these deviations is
discussed.

\section{Non-ideal discontinuity}
\label{Sec:IdealDipole} To model the finite-sized injectors we
used the current injection profile $\gamma_\theta(x)$ shown in
Fig.~\ref{Fig:dipole_layout}(d). Note that this shape is
a rather simple, but 
decent approximation, corresponding to our system with
superconducting injectors and a ground plane. 
Also, we consider here a 1D model, whereas in the real geometry
(see Fig.~\ref{Fig:sketch} or Fig. 3 of
Ref.~\onlinecite{Goldobin:Art-0-pi}) the injector current may
be distributed in a rather peculiar way across the top electrode.\\
Equation~(\ref{Eq:sG-phi}) can be rewritten as
\begin{equation}
  \phi_{xx}-\phi_{tt}-\sin\phi = \alpha\phi_t-\gamma-\gamma_\theta(x)
  , \label{Eq:sG-phi_gammatheta}
\end{equation}
where $\gamma_{\theta}(x)$ approximates $\theta_{xx}(x)$.
By introducing a finite injector size the $\kappa$-discontinuity
is now stretched over a finite
distance $\D l=2\Dw+\Dx$. 
In the injector area the dynamics of the Josephson phase and
therefore the junction properties are dominated by the injector
current distribution.

\begin{figure}[!tb]
  \centering \includegraphics* {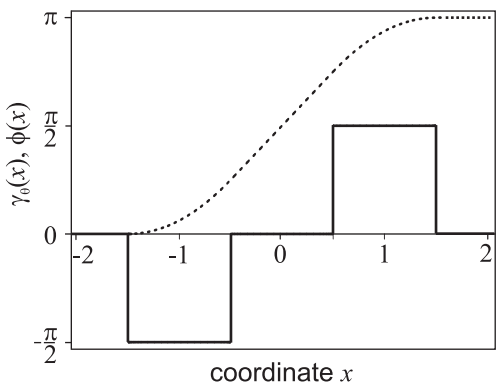}
  \caption{Spatial dependence of $\gamma_{\theta}(x)$ (solid line) and $\phi(x)$ (dotted line) in the injector area.
  $\Delta x = \Delta w =1$, $\gamma_{\rm inj} = -\pi/2$ for simplicity. The injector size is intentionally chosen to be huge
  ($\sim \lambda_J$) to visualize the behavior of the Josephson phase in all regions.}
  \label{Fig:gammatheta_x}
\end{figure}

First, we are interested in the static phase profile in the
injector area. If the injectors provide the current density
distribution $\gamma_{\theta}(x)$, the static sine-Gordon equation
is
\begin{equation}
\phi_{xx} = \sin \phi - \gamma -\gamma_\theta(x).
\label{Eq:Static-sG}
\end{equation}
For the following considerations it is more convenient to separate
the amplitude $\gamma_{\rm inj}$ of the injector current density
from the actual injection profile
\begin{equation}
\tilde{\gamma}_\theta(x)= \left\{
\begin{array}{rlccc}
 1&, &-\frac{\D l}{2}&< x<&-\frac{\Dx}{2}\\
 0&, &-\frac{\Dx}{2}&< x<&+\frac{\Dx}{2}\\
-1&, &+\frac{\Dx}{2}&< x<&+\frac{\D l}{2}
\end{array}
\right., \label{Eq:inj_profile}
\end{equation}
so that $\gamma_\theta(x) = \gamma_{\rm inj}
\tilde{\gamma}_\theta(x)$. In the injector region,
$\gamma_\theta(x)$ provides the dominating contribution
($|\gamma_{\rm inj}| \gg 1$) to the rhs. of
Eq.~(\ref{Eq:Static-sG}). If in addition $\Dx$ is assumed to be
$\ll 1$, Eq.~(\ref{Eq:Static-sG}) can be simplified to
\begin{equation}
\phi_{xx} = -\gamma_{\rm inj} \tilde{\gamma}_\theta(x).
\label{Eq:Static-sG-simple}
\end{equation}
Double integration of Eq.~(\ref{Eq:Static-sG-simple}) gives the
phase profile $\phi(x)$ in the region of injectors
($-\D{l}/2<x<+\D{l}/2$) %
%
depicted in Fig.~\ref{Fig:gammatheta_x}.%
%
The difference of the phase $\phi(\D{l}/2)-\phi(-\D{l}/2)$ is
proportional to $\gamma_{\rm inj}$ and should be equal to
$\kappa$. Thus we obtain a relation between $\kappa$ and the
amplitude of the injector current density $\gamma_{\rm inj}$
\begin{equation}
  \kappa = -\gamma_{\rm inj} \Dw(\Dw+\Dx)
  . \label{Eq:kappa(gamma_i)}
\end{equation}
\begin{figure}[!tb]
  \centering \includegraphics* {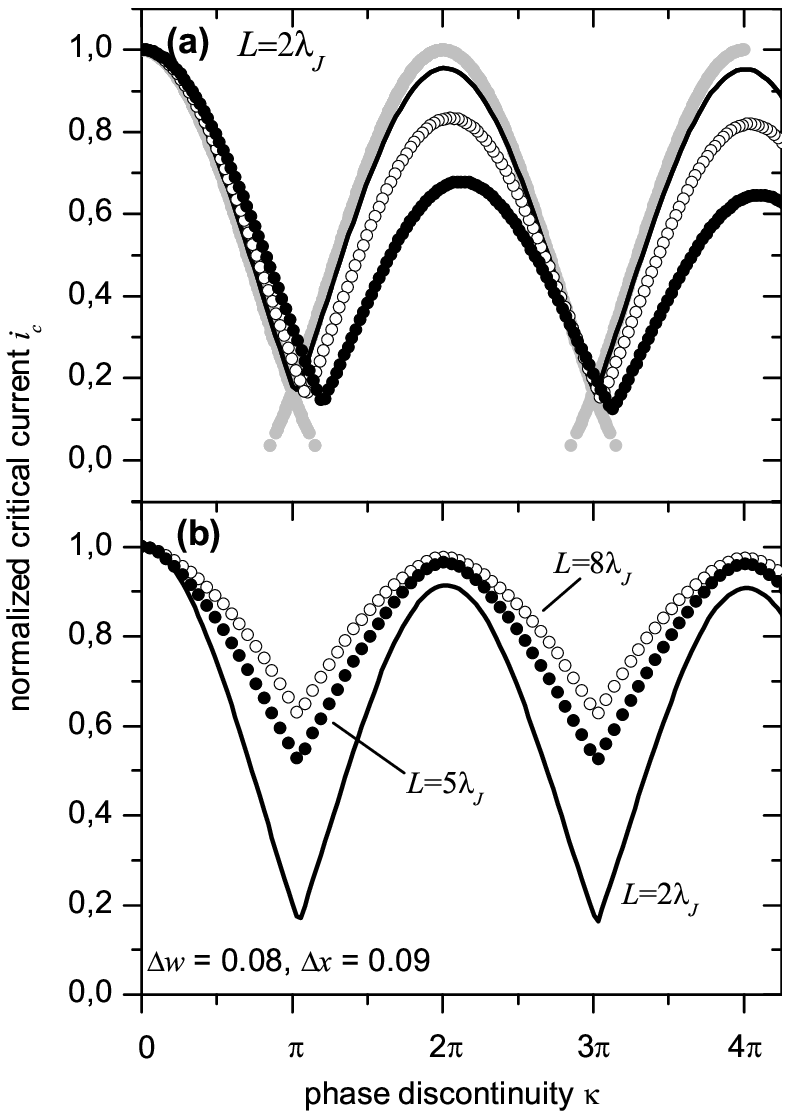}
  \caption{
      Normalized critical current $i_c$ \vs{} the Josephson phase discontinuity $\kappa$,
      numerically calculated for junctions with fixed $L = 2\lambda_J$ and different injectors
      (a) and junctions with same injectors ($\Delta w = 0.08$, $\Delta x =
      0.09$) but different $L$ (b). In (a) the sizes of injectors
      are $\Delta x = 0, \Delta w = 0$ (gray circles), $\Delta w = 0.04$, $\Delta x = 0.05$ (solid line), $\Delta w = 0.16$, $ \Delta x =0.17$ (open
      circles) and $\Delta w = 0.32$, $\Delta x = 0.33$ (black circles). }
  \label{Fig:Ic(inj)_flvsz}
\end{figure}

Now, let us consider the effect of a finite injector size on the
$i_c(\kappa)$ dependence. Fig.~\ref{Fig:Ic(inj)_flvsz} shows two
sets of numerically calculated $i_c(\kappa)$ curves. In
Fig.~\ref{Fig:Ic(inj)_flvsz}(a) junctions of the same length
$L=2\lambda_J$ but different injector sizes are compared, ranging
from $\Delta x = 0, \Delta w = 0$ (ideal discontinuity) to $\Delta
w = 0.32$, $\Delta x = 0.33$. The most prominent difference
compared to an ideal discontinuity
is a decrease in critical current at the side maxima $i_{c}(2\pi
n)$ ($n \neq 0$). The difference of critical currents
$i_c(0)-i_c(2 n \pi)$ becomes smaller as the injector size is
reduced and approaches zero as injectors become ideal ($\Delta x
\to 0, \Delta w \to 0$). In addition the minima of the curve shift
slightly towards values of $|\kappa| > \pi (2n+1)$ and $i_{c,{\rm
min}}$ is slightly reduced when injectors become very large and
comparable to the Josephson penetration depth. The last effect is
rather small and can normally be neglected for typical sizes of
injectors. Fig.~\ref{Fig:Ic(inj)_flvsz}(b) shows the $i_c(\kappa)$
dependence of junctions with the same injectors, but for different
LJJ lengths. Again, there is a decrease of $i_c(2 n \pi)$ for $n
\neq 0$; yet, with increasing junction length $L$ the
$i_c(\kappa)$ dependence approaches the curve of an ideal
discontinuity (see Fig.~\ref{Fig:Ic(kappa)}).
\begin{figure}[!tb]
  \centering \includegraphics*{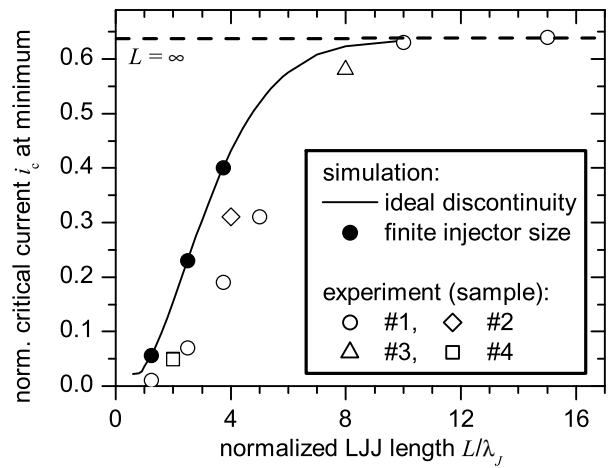}
  \caption{
      Minimum normalized critical current $i_c$ \vs junction
      length numerically calculated for an ideal discontinuity (solid line) and non-ideal
      injectors with $\Dx=0.21$, $\Dw=0.2$ (solid circles).
      Open symbols correspond to measurements of the samples summarized in Tab.~\ref{Tab:Samples}.
  }
  \label{Fig:Ic(pi)_length}
\end{figure}

Fig.~\ref{Fig:Ic(pi)_length} shows the numerically calculated
dependence of the minimum critical current $i_{c,{\rm min}}$ at
$\kappa \approx \pi$ on LJJ length $L$, for ideal and non-ideal
discontinuities, as well as experimentally obtained results. The
numerical calculations for the case of finite injector size were
performed with $\Delta w = 0.2$, $\Delta x = 0.21$, comparable to
the experimental dimensions of sample series~\#1. In the
simulations the finite injector size has a negligible influence on
$i_{c,{\rm min}}$. However, for $L < 10\,\lambda_J$ the
experimentally determined values of $i_{c,{\rm min}}$ are
systematically \emph{smaller} than theory predicts. Only for very
long JJs, $L \geq 10\,\lambda_J$, measurements and theory approach
the same asymptotic value of $2/\pi$ (semifluxon depinning
current~\cite{Kato:1997:QuTunnel0pi0JJ,Goldobin:SF-ReArrange,Zenchuk:2003:AnalXover}).
It is interesting to note, that the measured values (except for
sample \#2, that has smaller injectors) show a similar dependence
as in theory, yet the experimental data seem to be shifted along
the $L$-axis, towards longer JJs. So far this behavior is not
fully understood. However the better agreement between theory and
experimental data for sample \#2 gives already an indication, that
the actual injector size does play an important role and that the
1D model for the injectors could be too simple.

\begin{figure}[!tb]
  \centering \includegraphics* {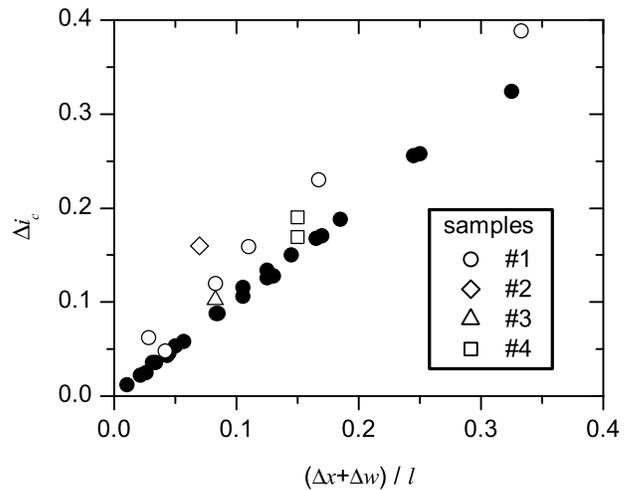}
  \caption{
      Numerically calculated (filled circles) and measured (open symbols) reduction of normalized critical current
      $\Delta i_c = 1 - i_c(2\pi)$
      for different junction lengths and injector sizes.
  }
  \label{Fig:Icdrop}
\end{figure}

Looking at the \emph{maxima} in Fig.~\ref{Fig:Ic(inj)_flvsz}
again, we conclude that the deviations from the ideal
characteristics depend on the relative size of the injector region
with respect to the junction length. There is a simple picture to
understand this qualitatively. For example, at $\kappa = 2\pi$ the
Josephson phase increases by $2\pi$, from $\phi(-\D l/2)$ to
$\phi(+\D l/2)$ in the region between the injectors, so that in
this region the net Josephson current across the junction can be
neglected (if the increase in $\phi$ is linear), no matter what
bias current is applied, because $\gamma \ll \gamma_{\rm inj}$.
Therefore the junction length $L$ is effectively reduced by the
size of the injector region and to a first order approximation the
critical current is $i_c(\pm 2\pi) \approx 1-(\Delta w + \Delta x
) / l$. Note, that the effective injector region, that has a net
Josephson current equal to zero, is only $\Dx + \Dw$ rather than
$\Dx + 2\Dw$, due to the parabolic shape of $\phi$ inside the
injector electrodes (cf. Fig.~\ref{Fig:gammatheta_x}). In
Fig.~\ref{Fig:Icdrop} we plot $\Delta i_c = 1- i_c(2\pi)$ \vs{}
$(\Delta w + \Delta x ) / l$ obtained experimentally and
numerically for junctions with different injector sizes and of
different LJJ lengths $l$. One can notice, that the experimental
data are in a good agreement with numerical calculations, yet the
measured drop in critical current
is always slightly larger than the theory predicts.\\
By confining even larger phase jumps ($\kappa = 4\pi, 6\pi,
\ldots$) into the area of injectors, the critical current of the
junction slowly decreases further (see
Fig.~\ref{Fig:Ic(inj)_flvsz}). Yet, the relative change $i_c(2\pi
n)- i_c(2\pi(n+1)),n > 0$ is much smaller than $\Delta
i_c=1-i_c(\pm 2\pi)$ and depends on the details of the phase
bending inside the injector area. This, however is a higher order
effect which is not addressed in this paper, because in
experiments one always wants to have $\Dx, \Dw \ll 1$.
%
%
\begin{figure}[!tb]
  \centering \includegraphics* {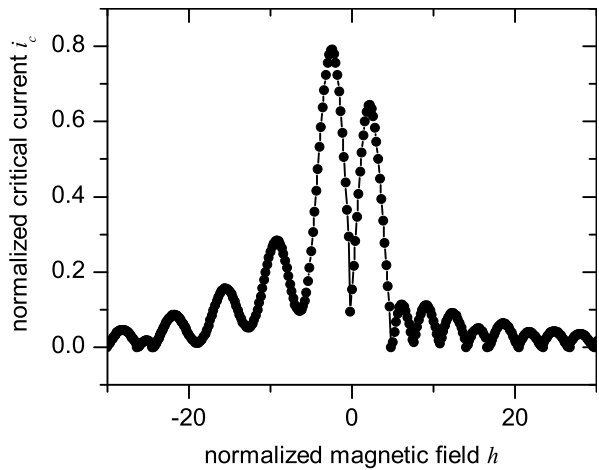}
  \caption{
      Numerically calculated $i_c(h)$ dependence of a LJJ ($L=2 \lambda_J$) in the 0-$\pi$-state.
      Injector size is $\Delta w = 0.16$ and $\Delta x = 0.13$. }
  \label{Fig:Ic(H)_real}
\end{figure}

\begin{figure}[!tb]
  \centering \includegraphics* {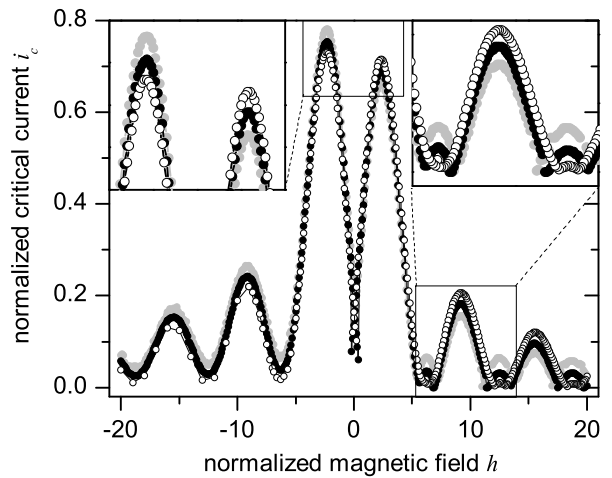}
  \caption{
      Numerically calculated $i_c(h)$ dependence of a LJJ ($L=2\lambda_J$) in the
      0-$\pi$-state. Injectors have a width/distance of $\Delta w =
      0.01 / \Delta x = 0.011$ (open circles), $\Delta w =
      0.04 / \Delta x = 0.05$ (black filled circles) and $\Delta w =
      0.08 / \Delta x = 0.09$ (gray filled circles).
  }
  \label{Fig:IcH_simmultiplot}
\end{figure}

Now let us concentrate on the 0-$\pi$-state [$I_{\rm inj}$
corresponding to a minimum in $I_c(I_{\rm inj})$] and consider the
$i_c(h)$ dependence, where $h=2H/H_{c1}$ and
$H_{c1}=\Phi_0/(2\pi\mu_0\lambda_L\lambda_J)$ is the first
critical field for a LJJ with thick electrodes. In
Fig.~\ref{Fig:Ic(H)_real} we show the numerically calculated
$i_c(h)$ dependence of an artificial 0-$\pi$-LJJ with finite
injectors. Parameters are chosen to be comparable with
experimental dimensions~\cite{Goldobin:Art-0-pi} (cf. sample \#4).
As has already been mentioned in
Ref.~\onlinecite{Goldobin:Art-0-pi}, simulated results reproduce
all features of the experimentally measured $I_c(H)$ dependence
rather well [see Fig.~\ref{Fig:Ic(H)_measured}(b) or Fig.~5 in
Ref.~\onlinecite{Goldobin:Art-0-pi}]. Although the curve shown in
Fig.~\ref{Fig:Ic(H)_real} has the characteristic minimum of the
critical current in zero magnetic field, it is rather asymmetric.
The maximum critical currents of the side lobes are different and
the periodicities of the outer oscillations for negative and
positive field differ. Yet, with decreasing injector size this
asymmetry disappears as can be seen in
Fig.~\ref{Fig:IcH_simmultiplot}.
%


To understand the reason for this asymmetries, we, for the moment,
consider a relatively short JJ ($L \lesssim \lambda_J$) and find
the expression for $i_c(h)$ for injectors having a normalized size
$\Dw$ and distance $\Dx$ between them. The Josephson phase is
assumed to be linear outside the injector area, \ie,
%
\begin{equation}
    \phi_{out}(x)=
    \left\{
      \begin{array}{rcrcl}
        \phi_0+h\left( x+ \frac{\Delta l }{2}  \right)
        &,&  -\frac{l}{2}<&x&<-\frac{\Delta l}{2}\\
        \phi_0+\kappa+h\left( x-\frac{\Delta l}{2}  \right)
        &,& +\frac{\Delta l}{2}<&x&<+\frac{l}{2}
      \end{array}
    \right.
  . \label{Eq:PhaseOutside}
\end{equation}
%
Inside the injector region the phase changes from $\phi_0$ to
$\phi_0+\kappa$ over the distance $\Delta l=2\D w +\D x$, but its
particular profile will be discussed later.

The total supercurrent carried by the LJJ outside the injector
area is calculated by integrating $\sin\phi_{out}(x)$ over the
junction length excluding the injector region. By varying $\phi_0$
to maximize $i_c$ we find that the maximum $i_c$ is reached at
\begin{equation}
  \phi_0=\frac{\pi}{2}-\frac{\kappa}{2}
  , \label{Eq:phi0}
\end{equation}
and is given by
\begin{equation}
  i_c(h) = \frac{4}{hl}\cos\frac{\kappa+h(l-\D l)}{2}\sin\frac{h(l-\D l)}{2}
  . \label{Eq:gamma_c(h)}
\end{equation}
The dependence (\ref{Eq:gamma_c(h)}) is an odd function of $h$,
\ie, it is anti-symmetric with respect to the origin $h=0$. We
note that in experiment one always measures the absolute value
$|i_c(h)|$, which is a \emph{symmetric} (even) function with
respect to the origin $h=0$, and has a minimum at $h=0$.

Now let us calculate the supercurrent carried by the region of
injectors. The phase $\phi(x)$ changes from $\phi_0$ to
$\phi_0+\kappa$ as shown in Fig.~\ref{Fig:gammatheta_x}, \ie, it
has two parabolic parts corresponding to two injectors and a
linear part corresponding to the spacing between injectors. To
calculate the supercurrent $i_{s,{\rm inj}}$ carried by the
injector region, we again integrate
\begin{equation}
  i_{s,{\rm inj}} = \frac{1}{l} \int_{-\Delta l/2}^{+\Delta l/2}\sin\phi(x)\,dx
  . \label{Eq:i_s}
\end{equation}
Since some parts of $\phi(x)$ are parabolic, we will end up with
Fresnel integrals which will only add additional complications. We
thus change the phase profile a little bit to simplify
integration, taking into account that the injector current profile
shown in Fig.~\ref{Fig:dipole_layout} is only a
simple approximation.
 Considering that in our original model $\phi(x)$ changes
parabolically, \ie, slowly, and that for superconducting injectors
most of the current flows close to the inner edge of the injector,
we adopt a very simple injector phase profile
\begin{widetext}
\begin{equation}
  \phi(x) = \phi_0 + \kappa\left( \frac{x}{\Dx+\Dw} + \frac12\right)
  ,\quad -\frac{\Dx+\Dw}{2}<x<+\frac{\Dx+\Dw}{2}, \label{Eq:phi_i}
\end{equation}
\end{widetext}
\ie, the phase changes linearly from $\phi_0$ to $\phi_0+\kappa$
only in the space between the injector centers. By effectively
shortening the injector area, the still unattended outer halves of
the injectors are simply added to the part of the LJJ without
injectors, \ie, we substitute $\Delta \tilde{l} = \Delta w +
\Delta x$ for $\D l$ in Eqs.~(\ref{Eq:PhaseOutside}),
(\ref{Eq:gamma_c(h)}) and (\ref{Eq:i_s}). Thus, the supercurrent
(\ref{Eq:i_s}) carried by the injector region calculated for
optimum $\phi_0$ (\ref{Eq:phi0}) is
\begin{equation}
  i_{s,{\rm inj}} = (\Dx+\Dw)\frac{2}{\kappa l}\sin\frac{\kappa}{2}
  . \label{Eq:i_s-approx}
\end{equation}
Note, that $i_{s,{\rm inj}}$ does not depend on magnetic field,
because we did not take the penetration of magnetic field into the
injector region into account.

\begin{figure*}[!tb]
  \centering \includegraphics* {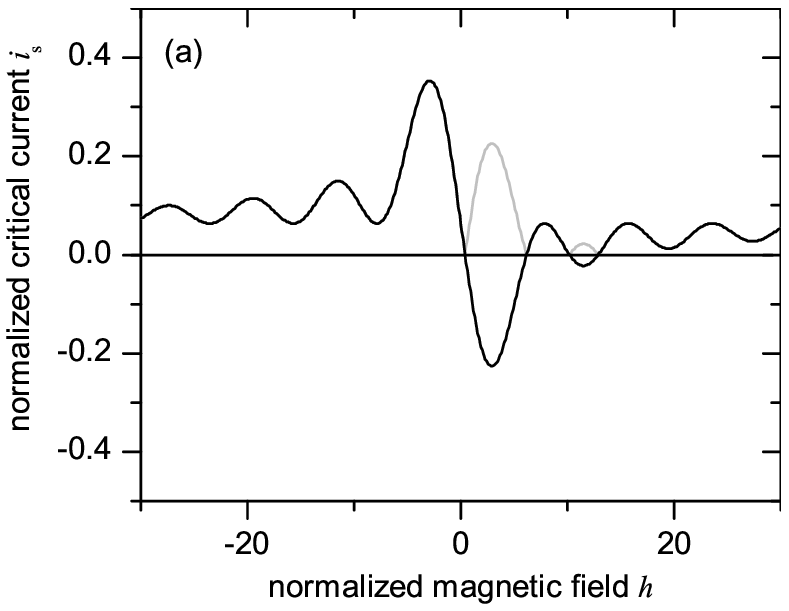}
  \centering \includegraphics* {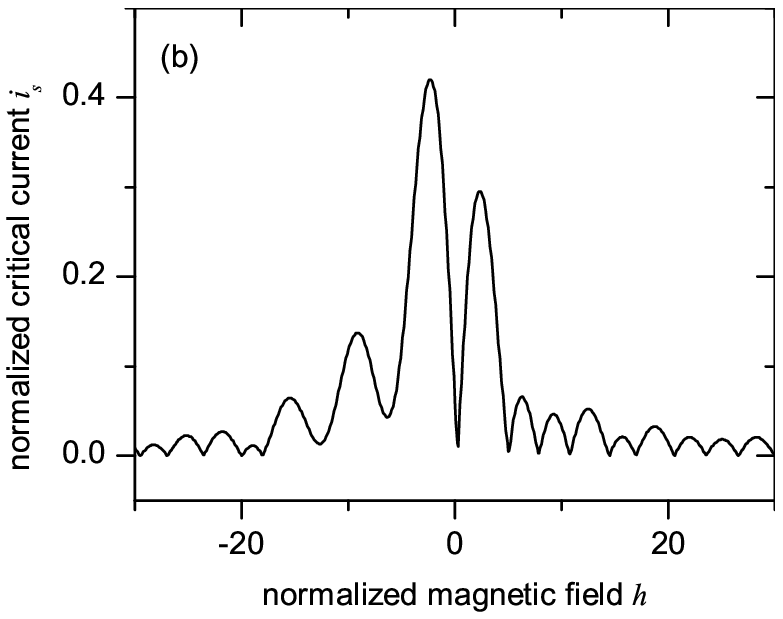}
  \caption{(a): Calculated dependence of $i_s(h)$ (black line) and $|i_s(h)|$ (gray line).
  (b): Calculated dependence of $|i_s(h)|$ in case magnetic field is penetrating the injector area.
  Parameters are $l = 2$, $\kappa = \pi$ and $\Dx = \Dw = 0.2$.}
  \label{Fig:Ic(H)_maplesim}
\end{figure*}

The resulting normalized maximum supercurrent \vs{} field
dependence which one measures in experiment is given by
\begin{equation}
  i_s(h) = |i_c(h)+i_{s,{\rm inj}}|
  , \label{Eq:i_c_tot}
\end{equation}
\ie, the anti-symmetric $i_c(h)$ dependence from
Eq.~(\ref{Eq:gamma_c(h)}) is first shifted along the $i$-axis by
$i_{s,{\rm inj}}$ and then the absolute value is calculated, [see
Fig.~\ref{Fig:Ic(H)_maplesim}~(a)].
As a result one gets asymmetric maxima for $h>0$ and $h<0$.
Moreover, the side oscillations of $i_c(h)+i_{s,{\rm inj}}$ due to
the shift may not intersect the $h$-axis anymore for $h<0$, but do
intersect it for $h>0$, resulting in very different patterns of
the side maxima for the $i_s(h)$ dependence, [see
Fig.~\ref{Fig:Ic(H)_maplesim}(a)].

Before we proceed with a comparison of experiment with our
analytical results, we would like to note another effect which may
become noticeable in not very long junctions with rather large
injectors. Namely, the phase jump $\kappa$ which, up to now, was
supposed to be only due to the injector current
$\kappa=\kappa_{\rm inj}$, can also have a contribution from the
magnetic field penetrating the injector area. We write
$\kappa=\kappa_{\rm inj}+h\,(\Dx+\Dw)$ and use the so-defined
$\kappa$ in Eq.~(\ref{Eq:i_s-approx}) and (\ref{Eq:i_c_tot}).
Then, the contribution of the injector area to the supercurrent,
\ie, the above mentioned shift, becomes field dependent. This
means that side oscillations of $i_c(h)+i_{s,{\rm inj}}$ which did
not cross the ($i_s=0$)-axis for lower fields, may do so for
larger fields, as is shown in Fig.~\ref{Fig:Ic(H)_maplesim}(b).
The larger the ratio $(\Dx+\Dw)/l$ the more noticeable this effect
will be. Fig.~\ref{Fig:Ic(H)_measured} shows experimentally
obtained $I_c(H)$-dependences of two LJJ of different length in
the $0$-$\pi$-state. As one can see clearly, these measurements
are in good agreement with the above calculations. In particular,
one can notice that the shorter the JJ is, the more pronounced is
the asymmetry in $i_c(h)$.

It is interesting to note, that with the simple short junction
model presented above, almost all features of the measured
$I_c(H)$ dependence can be explained and effectively pinpointed to
the supercurrent inside the injector area.

\section{Conclusions}
\label{Sec:Conclusion}

A pair of tiny current injectors can be used to create an
arbitrary phase jump of the Josephson phase and to study arbitrary
fractional vortices pinned at it. An important step, which one
should make following this approach, is to calibrate the
injectors, \ie, to measure the $I_c(I_{\rm inj})$ dependence. Here
we calculated how $I_c(I_{\rm inj})$ should look like in the
$0$-$\kappa$-LJJ of length $L$ with an ideal discontinuity. We
discuss the effect of finite injector size on the $I_c(I_{\rm
inj})$ and $I_c(H)$ dependence and find rather good agreement
between our model and experimental data. The consequences of
finite sized injectors compared to an ideal discontinuity can be
summarized as follows:

(I) The $i_c(\kappa)$ dependence is not perfectly $2\pi$ periodic
but exhibits a decrease in $i_c(\pm 2n\pi)<1$, $n \neq 0$. The
drop of $i_{c}$ at $\kappa = \pm 2\pi$ can roughly be estimated as
$1-i_c(\pm 2\pi) \approx (\Delta w + \Delta x)/l$ and is
associated with the exclusion of injector area that carries no net
supercurrent.

(II) The $i_c(h)$ characteristic for $\kappa=\pi$ is asymmetric
with respect to magnetic field. This asymmetry is caused by the
Josephson currents flowing in the area between the injectors.
These Josephson currents shift the $i_c(h)$ curve by $i_{s,{\rm
inj}}$ [cf. Eq.~(\ref{Eq:i_s-approx})] and become weakly field
dependent for $\D l \lesssim l$. As in experiment only
$i_s(h)=|i_c(h)+i_{s,{\rm inj}}|$ can be measured, yet $i_c(h)$ is
an odd function of $h$, parts of the $i_c(h)$ curve get reflected
at the $h$-axis, resulting in a field dependent and asymmetric
pattern of the higher order maxima. The difference in normalized
critical current between the main maxima is $\approx \pm
\frac{4}{\pi}(\Delta x +
\Delta w)/l$ with respect to ideal injectors.\\
In general, it is desirable to make the injector region as small
as possible, to avoid the discussed finite size effects. However,
for cases where this is not possible, our results allow one to
interpret the experimental data properly and optimize the design
of such LJJs. 

\begin{acknowledgments}
  This work was supported by the Deutsche Forschungsgemeinschaft (project GO-1106/1) and by the ESF program "PiShift".
\end{acknowledgments}

\bibliography{LJJ,pi,SF,SFS,QuComp,software}
\co{0}{
  \section*{unplugged}
  \begin{itemize}
    \item Fine structure and resonances. Compare 0-$\pi$-LJJs of different $L$.
    \item Address decay of $I_{c,max}(I_{\rm inj})$.
    \item Address hysteresis at $I_c(I_{\rm inj})$ at large $I_{\rm inj}$.
    \item Address small shift of $I_c(I_{\rm inj})$
    \item Get $I_c(I_{\rm inj})$ pattern for large $L$
    \item Present $I_{\max}(H)$ and $I_{\max}(I_{\rm inj})$ for semi-integer ZFS.
  \end{itemize}
}

\end{document}